\documentclass[preprin,preprintnumbers,eqsecnum,amsmath,amssymb,nofootinbib]{revtex4}
\usepackage{graphics,epsfig}
\usepackage{graphicx}
\usepackage{dcolumn}
\usepackage{bm}

\begin{document}
\baselineskip=0.8 cm
\title{ Rotating BTZ-like black hole and central charges in Einstein-bumblebee gravity}

\author{Chikun Ding$^{1,2}$}\email{dingchikun@163.com ; dck@hhtc.edu.cn}
\author{Yu Shi$^{1}$}\author{Jun Chen$^{1}$}\author{Yuebing Zhou$^{1}$}\author{Changqing Liu$^{1}$}\author{Yuehua Xiao$^{1}$}
\affiliation{$^1$Department of Physics, Huaihua University, Huaihua, 418008, P. R. China\\
$^2$Key Laboratory of Low Dimensional
Quantum Structures and Quantum Control of Ministry of Education,
and Synergetic Innovation Center for Quantum Effects and Applications,
Hunan Normal University, Changsha, Hunan 410081, P. R. China}

\vspace*{0.2cm}
\begin{abstract}
\baselineskip=0.6 cm
\begin{center}
{\bf Abstract}
\end{center}

We obtain an exact rotating BTZ-like black hole solution by solving the corresponding gravitational field equations and the bumblebee motion equations in Einstein-bumblebee gravity theory. Result is presented for the purely radial Lorentz symmetry violating and can only exist with a linear functional potential of the bumblebee field. This black hole has two horizons and an ergosphere which are dependent on the bumblebee coupling constant $\ell$. The concepts of the area and volume of the horizon should be renewed in this LV spacetime due to the nontrivial contribution of coupling between the bumblebee field and the Ricci tensor. Only in this way, the entropy-area relation, first law of thermodynamics and the Smarr formula can still be constructed. We also study the AdS/CFT correspondence of this black hole, find that the entropy product of its inner and outer horizons is universal. So the central charges of the dual CFT on the boundary can be obtained via the thermodynamic method, and they can reappear black hole mass and angular momentum in the bulk.

\end{abstract}

 \maketitle

\vspace*{0.2cm}
\section{Introduction}

The standard model (SM) of particle physics and the general relativity (GR) cannot explain everything in the universe, such as dark energy and that what occurs in the vicinity of a black hole. Thus, on the very high energy scales, one reconsiders combining the SM with GR in a unified theory, i.e., "quantum gravity". The standard model extension (SME), proposed by Kostelesk\'{y} and collaborators \cite{Kostelecky1,Kostelecky2,Kostelecky3,Kostelecky4,Kostelecky5,Kostelecky6,Kostelecky7,Kostelecky8}, is this effective field theory combining GR and SM at low energy scales, and incorporates all possible background fields that violate the fundamental symmetries existent in nature, the Lorentz invariance, happening on the high energy scales (about Planck scale). Studying the Lorentz violation (LV) is a useful approach toward investigating the foundations of modern physics. Besides SME, string theory \cite{kostelecky198939}, noncommutative field theories \cite{carroll,carroll2,carroll3}, spacetime-varying fields \cite{bertolami69,bertolami692,bertolami693}, loop quantum gravity theory \cite{gambini,gambini2}, brane world scenarios \cite{burgess03,burgess032,burgess033}, massive gravity \cite{fernando} and Einstein-aether theory \cite{jacobson,jacobson2} are other proposals of Lorentz violation.

SME can be used to calculate a number of predictions which can be tested in modern high-precision experiments \cite{Kostelecky8,tasson}. The primary LV in the gravity sector of SME is the form of $s^{\mu\nu}R_{\mu\nu}$, where $s^{\mu\nu}$ is a tensor field possessing a nonzero background configuration and defines preferred frames, $R_{\mu\nu}$ is the Ricci tensor. The simplest form of $s^{\mu\nu}$ is the bumblebee field $B^\mu$, which is expected to produce new and maybe even eccentric phenomena through the coupling term $B^{\mu}B^{\nu}R_{\mu\nu}$. The name of "bumblebee" comes from the alleged remark that bumblebees should not be allowed to fly since we did not know how their wings produce lift \cite{uniyal}. The surprising property of this bumblebee gravity is that, unlike the absence of $U(1)$ gauge symmetry, it does not forbid the propagation of massless vector modes \cite{bluhm2008}. So one expects to reveal a variety of physical relics which may be of interest in studies of dark energy and dark matter due to the appearance of Nambu-Goldstone and massive Higgs in theories with the spontaneous Lorentz symmetry breaking \cite{Kostelecky6,Kostelecky7,bluhm}.

In this Einstein-bumblebee gravity theory, Bertolami and P\'aramos studied the 4-dimensional static vacuum solutions including the purely radial, the radial/temporal or the axial/temporal LV \cite{bertolami}. They found that there exists an exact black solution for the purely radial bumblebee field; for the radial/temporal LV, there exists only a slightly perturbed metric where one cannot constrain the physical parameters from the observed limits on the PPN(parameterized post-Newtonian) parameters.  In recent years, Casana {\it et al} \cite{casana} obtained an exact Schwarzschild-like black hole solution for the purely radial bumblebee field. Xu {\it et al} studied the radial/temporal bumblebee field and the properties of some general numerical static black hole solutions \cite{xu}. In 2020, we obtained an exact Kerr-like solution and studied its black hole shadow \cite{ding2020}. Though this solution does not seem to satisfy the bumblebee field equations, Liu {\it et al} found that it can still be satisfied under some certain conditions, i.e., considered as an approximate solution of the bumblebee field equations \cite{liu}. After that we derived a slowly rotating black hole solution which satisfies all field equations \cite{ding2021}. Lately, we derived a black hole solution and a cosmological solution of Einstein-Gauss-Bonnet gravity coupled to bumblebee field \cite{ding2022}, found that the Guass-Bonnet term and the bumblebee field can both act as a form of dark energy. In  a high dimensional spacetime, we obtained an exact  AdS-like black hole solution \cite{ding2023}, found that the conceptions of black hole horizon area/entropy and volume inside horizon should be renewed due to its anisotropy.

In this paper, we would seek a rotating black hole solution in 3-dimensional spacetime. The first 3-dimensional black hole solution is BTZ(Ba\~{n}ados, Teitelboim and Zanelli) solution \cite{banados1,banados2}, which is asymptotically anti-de Sitter(AdS) and has on curvature singularity. It is a genuine black hole solution due to the presence of event horizon, Hawking radiation, entropy,  and playing a significant role to understand physical properties in higher dimensions by using many of toy models \cite{bengtsson1,bengtsson2}.  The black hole thermodynamics has raised some challenging questions: a statistical derivation of black hole entropy and an account of its microstates. A promised idea is the holographic principle: there is a nontrivial match between features of 2-dimensional conformal field theory (CFT) and features of black holes \cite{castro}. The microscopic degrees of freedom of the black hole are described by CFT living on the boundary. Quantum studies around BTZ black holes can help better understand AdS/CFT correspondence, T-duality and U-duality to classes of asymptotically flat black strings \cite{horowitz1,horowitz2}.

We will derive a 3-dimensional  rotating black hole solution and study the central charges of the dual CFT in the theory of Einstein gravity coupled to the bumblebee fields. The rest paper is organized as follows.   In Sec. II we give the background for the Einstein-bumblebee theory. In Sec. III, we give the  black hole solution by solving the gravitational and bumblebee field equations. In Sec. IV, we study its thermodynamics and central charges  and find some effects of the Lorentz breaking constant $\ell$. Sec. V is for a summary.

\section{Einstein-bumblebee gravity in 3-dimensional spacetime}

In the bumblebee gravity model, one introduces the bumblebee vector field $B_{\mu}$ which has a nonzero vacuum expectation value, to lead a spontaneous Lorentz symmetry breaking in the gravitational sector via a given potential. In the three dimensional spacetime, the
 action of Einstein-bumblebee gravity is \cite{ding2021},
\begin{eqnarray}
\mathcal{S}=
\int d^3x\sqrt{-g}\Big[\frac{R-2\Lambda}{2\kappa}+\frac{\varrho}{2\kappa} B^{\mu}B^{\nu}R_{\mu\nu}-\frac{1}{4}B^{\mu\nu}B_{\mu\nu}
-V(B_\mu B^{\mu}\pm b^2)+\mathcal{L}_M\Big], \label{action}
\end{eqnarray}
where $R$ is Ricci scalar and $\Lambda$ is the negative cosmological constant. $\kappa=4\pi G/c^4$ for the three dimensions, where $G$ is the Newtonian constant.

The coupling constant $\varrho$ dominates the non-minimal gravity interaction to bumblebee field $B_\mu$. The term $\mathcal{L}_M$ represents possible interactions with matter or external currents.
The constant $b$ is a real positive constant. The potential $V(B_\mu B^{\mu}\pm b^2)$ triggers Lorentz and/or $CPT$ (charge, parity and time) violation. It gives a nonzero vacuum expectation value (VEV) for bumblebee field $B_{\mu}$, $\langle B^{\mu}\rangle= b^{\mu}$, indicating that the vacuum of this model obtains a prior direction in the spacetime.
The two classes of the potentials have been investigated in Ref. \cite{bluhm2008}, i.e., the smooth functionals
and Lagrange-multiplier functionals.
Another vector $b^{\mu}$ is a function of the spacetime coordinates and has a constant value $b_{\mu}b^{\mu}=\mp b^2$, where $\pm$ signs mean that $b^{\mu}$ is timelike or spacelike, respectively.
The bumblebee field strength is
\begin{eqnarray}
B_{\mu\nu}=\partial_{\mu}B_{\nu}-\partial_{\nu}B_{\mu}.
\end{eqnarray}
This antisymmetry of $B_{\mu\nu}$ implies the constraint \cite{bluhm}
\begin{eqnarray}
\nabla ^\mu\nabla^\nu B_{\mu\nu}=0.
\end{eqnarray}

Varying the action (\ref{action}) with respect to the metric yields the gravitational field equations
\begin{eqnarray}\label{einstein0}
G_{\mu\nu}+\Lambda g_{\mu\nu}=\kappa T_{\mu\nu}^B+\kappa T_{\mu\nu}^M,
\end{eqnarray}
where $G_{\mu\nu}=R_{\mu\nu}-g_{\mu\nu}R/2$, and the bumblebee energy momentum tensor $T_{\mu\nu}^B$ is
\begin{eqnarray}\label{momentum}
&&T_{\mu\nu}^B=B_{\mu\alpha}B^{\alpha}_{\;\nu}-\frac{1}{4}g_{\mu\nu} B^{\alpha\beta}B_{\alpha\beta}- g_{\mu\nu}V+
2B_{\mu}B_{\nu}V'\nonumber\\
&&+\frac{\varrho}{\kappa}\Big[\frac{1}{2}g_{\mu\nu}B^{\alpha}B^{\beta}R_{\alpha\beta}
-B_{\mu}B^{\alpha}R_{\alpha\nu}-B_{\nu}B^{\alpha}R_{\alpha\mu}\nonumber\\
&&+\frac{1}{2}\nabla_{\alpha}\nabla_{\mu}(B^{\alpha}B_{\nu})
+\frac{1}{2}\nabla_{\alpha}\nabla_{\nu}(B^{\alpha}B_{\mu})
-\frac{1}{2}\nabla^2(B_{\mu}B_{\nu})-\frac{1}{2}
g_{\mu\nu}\nabla_{\alpha}\nabla_{\beta}(B^{\alpha}B^{\beta})\Big].
\end{eqnarray}
The prime denotes differentiation with respect to the argument,
\begin{eqnarray}
V'=\frac{\partial V(x)}{\partial x}\Big|_{x=B^{\mu}B_{\mu}\pm b^2}.
\end{eqnarray}
Varying instead with respect to the the bumblebee field generates the bumblebee equations of motion (supposing that there is no coupling between the bumblebee field and $\mathcal{L}_M$),
\begin{eqnarray}\label{motion}
\nabla ^{\mu}B_{\mu\nu}=2V'B_\nu-\frac{\varrho}{\kappa}B^{\mu}R_{\mu\nu}.
\end{eqnarray}
The contracted Bianchi identities ($\nabla ^\mu G_{\mu\nu}=0$) lead to conservation of the total energy-momentum tensor
\begin{eqnarray}\label{}
\nabla ^\mu T_{\mu\nu}=\nabla ^\mu\big( T^B_{\mu\nu}+T^M_{\mu\nu}\big)=0.
\end{eqnarray}

We suppose that there is no matter field and the bumblebee field is frosted at its VEV like in Refs \cite{casana,bertolami}, i.e., it is
\begin{eqnarray}\label{bbu}
B_\mu=b_\mu.
\end{eqnarray}
And the potential has a smooth quadratic function or a linear Lagrange-multiplier function form \cite{bluhm2008} \begin{eqnarray}
V=\frac{k}{2}x^2;\qquad V=\frac{\lambda}{2}x,\qquad x=(B_\mu B^\mu-b^2),
\end{eqnarray}
where $k$ is a constants and $\lambda$ is a Lagrange-multiplier field which is auxiliary and has no kinetic terms \footnote{The form of Lagrange-multiplier potential $V=\lambda(B_\mu B^\mu-b^2)/2$ is similar to the constrain term $\lambda(u^2+1)$ in Einstein-aether gravity theory, where $u^2=u_au^a$, and $u^a$ is the aether field (cf. the last term of Eq. 2.3 in Ref. \cite{ding2016}).}.
The both potentials are $V=0$ under the condition (\ref{bbu}).  Then the first three terms in Eq. (\ref{momentum}) are like those of the electromagnetic field, the distinctive features are the coupling items to Ricci tensor and the first order derivative of the potential $V'$. In four and higher dimensional spacetime \cite{maluf,ding2023}, by applying the smooth potential $V=kx^2/2$, the authors found that there is non black hole solution with nonzero cosmological constant $\Lambda\neq0$.
Under this condition,  Eq. (\ref{einstein0}) leads to gravitational field equations \cite{ding2021}
\begin{eqnarray}\label{bar}
G_{\mu\nu}+\Lambda g_{\mu\nu}=\kappa (2V' b_\mu b_\nu+b_{\mu\alpha}b^{\alpha}_{\;\nu}-\frac{1}{4}g_{\mu\nu} b^{\alpha\beta}b_{\alpha\beta})+\varrho\Big(\frac{1}{2}
g_{\mu\nu}b^{\alpha}b^{\beta}R_{\alpha\beta}- b_{\mu}b^{\alpha}R_{\alpha\nu}
-b_{\nu}b^{\alpha}R_{\alpha\mu}\Big)
+\bar B_{\mu\nu},
\end{eqnarray}
with
\begin{eqnarray}\label{barb}
&&\bar B_{\mu\nu}=\frac{\varrho}{2}\Big[
\nabla_{\alpha}\nabla_{\mu}(b^{\alpha}b_{\nu})
+\nabla_{\alpha}\nabla_{\nu}(b^{\alpha}b_{\mu})
-\nabla^2(b_{\mu}b_{\nu})-g_{\mu\nu}\nabla_\alpha\nabla_\beta(b^\alpha b^\beta)\Big].
\end{eqnarray}
\section{Rotating BTZ-like black hole solution}
The stationary axial symmetric black hole metric in the three dimensional spacetime  has the form
\begin{eqnarray}\label{metric}
&&ds^2=-e^{2\Phi(r)}dt^2+e^{2\psi(r)}dr^2+r^2\big[\Omega(r) dt+d\phi\big]^2,
\end{eqnarray}
where $\Phi(r), \psi(r)$ and $\Omega(r)$ are some undetermined functions.
In the present study, we pay attention to that the bumblebee field has a radial vacuum energy expectation because that the spacetime curvature has a strong radial variation, on the contrary that the temporal changes are very slow. So the bumblebee field is supposed to be spacelike($b_\mu b^\mu=$ positive constant) as that
\begin{eqnarray}\label{bu}
b_\mu=\big(0,b_0e^{\psi(r)},0\big),
\end{eqnarray}
where $b_0$ is a positive constant.
Then the bumblebee field strength is
\begin{eqnarray}
b_{\mu\nu}=\partial_{\mu}b_{\nu}-\partial_{\nu}b_{\mu},
\end{eqnarray}
whose components are all zero. And their divergences are all zero, i.e.,
\begin{eqnarray}
\nabla^{\mu}b_{\mu\nu}=0.
\end{eqnarray}
From the equation of motion (\ref{motion}), we have\footnote{If the coupling constant $\varrho=0$, then it leads $V'=0$, and the bumblebee energy momentum tensor $T_{\mu\nu}^B=0$ which will result in BTZ black hole solution.}
\begin{eqnarray}
b^{\mu}R_{\mu\nu}=\frac{2\kappa}{\varrho}V'\label{motion2}.
\end{eqnarray}
The gravitational field equations (\ref{bar}) become
\begin{eqnarray}\label{}
G_{\mu\nu}+\Lambda g_{\mu\nu}=\bar B_{\mu\nu}+2\kappa V'b_\mu b_\nu+\varrho\Big(\frac{1}{2}
g_{\mu\nu}b_0^2e^{-2\psi}R_{11}- b_{\mu}b^{\alpha}R_{\alpha\nu}
-b_{\nu}b^{\alpha}R_{\alpha\mu}\Big).
\end{eqnarray}

 For the metric (\ref{metric}), the nonzero components of Einstein tensor $G_{\mu\nu}$, Ricci tensor $R_{11}$ and the bumblebee tensor $\bar B_{\mu\nu}$ are shown in the appendix A.
Substituting the equation $G_{22}+\Lambda g_{22}=\bar B_{22}$ into $G_{02}+\Lambda g_{02}=\bar B_{02}$, one can obtain
\begin{eqnarray}\label{2202}
(1+\ell)\big[(r\Omega''+3\Omega')-r\Omega'(\Phi'+\psi')\big]=0,
\end{eqnarray}
where $\ell=\varrho b_0^2$ and the symbol $\prime$ means the derivative with their argument.
By using the motion equation (\ref{motion2}), one can obtain
\begin{eqnarray}\label{r11}
R_{11}=\frac{2\kappa V'}{\varrho}e^{2\psi}.\end{eqnarray}
Substituting the equations $G_{22}+\Lambda g_{22}=\bar B_{22}$ and (\ref{r11}) into $G_{00}+\Lambda g_{00}=\bar B_{00}$, one can obtain
\begin{eqnarray}\label{2200}
(1+\ell)\left[r\Omega(r\Omega''+3\Omega')+\frac{1}{4}r^2\Omega'^2
-e^{2\Phi}\frac{\psi'}{r}-r^2\Omega\Omega'(\Phi'+\psi')\right]+e^{2\Phi+2\psi}\Lambda=0.
\end{eqnarray}
Substituting the equation (\ref{r11}) into $G_{11}+\Lambda g_{11}=\bar B_{11}$, one can obtain
\begin{eqnarray}\label{r11rr}
(1+\ell)\left[e^{2\Phi}\frac{\Phi'}{r}+\frac{1}{4}r^2\Omega'^2\right]+e^{2\Phi+2\psi}\Lambda=0.
\end{eqnarray}
The above two equations (\ref{2200}) and (\ref{r11rr}) can give that
\begin{eqnarray}\label{1100}
r\Omega\big[(r\Omega''+3\Omega')-r\Omega'(\Phi'+\psi')\big]-\frac{1}{r}e^{2\Phi}(\Phi'+\psi')=0.
\end{eqnarray}
Then the Eqs. (\ref{2202}) and (\ref{1100}) can give the following equations
\begin{eqnarray}
&&\Phi'+\psi'=0,\label{pp}\\&&r\Omega''+3\Omega'=0\label{omega}.
\end{eqnarray}
Eq. (\ref{omega}) can give
\begin{eqnarray}
\Omega=-\frac{j}{2r^2},
\end{eqnarray}
where $j$ is an integral constant relating to its angular momentum.
From the Eq. (\ref{pp}), one can assume that $e^{2\Phi}=f(r)$ and  $e^{2\psi}=C/f(r)$, where $C$ is a constant to be determined. Substituting it into the equations $G_{11}+\Lambda g_{11}=\bar B_{11}$ and $G_{22}+\Lambda g_{22}=\bar B_{22}$, one can obtain that
\begin{eqnarray}
\big(rf''+f'-\frac{j^2}{r^3}\big)+\frac{4C}{1+\ell}\Lambda r=0,
\end{eqnarray}
which can give that
\begin{eqnarray}\label{metricf}
f(r)= -m-\frac{C\Lambda r^2}{(1+\ell)}+\frac{j^2}{4r^2},
\end{eqnarray}
where $m$ is also an integral constant relating to its mass.
In order to get a BTZ-like solution, we choose the constant $C=(1+\ell)$.

\subsection{Non black hole solution for $V=kx^2/2$}
Though the bumblebee field motion equation (\ref{r11}) is used in the above derivations, the solution (\ref{metricf}) would  be uncertain to satisfy it.
The bumblebee field motion equation (\ref{r11}) can be rewritten as following
\begin{eqnarray}\label{motion3}
\frac{f'}{r}-f''+\frac{j^2}{4r^2}= \frac{4\kappa V'C}{\varrho}.
\end{eqnarray}
 If the bumblebee potential takes the form of $V=kx^2/2$, then $V'=0$ at its VEV. Substituting the solution (\ref{metricf}) into the above Eq. (\ref{motion3}), one can obtain the zero cosmologic constant, i.e., $\Lambda=0$. Therefore, in this case, there is no black hole solution indeed with $\Lambda\neq0$. It is consistent with the four and higher dimensions \cite{maluf,ding2023}.

 \subsection{New black hole solution for $V=\lambda x/2$}

 If the bumblebee potential takes the form of $V=\lambda x/2$, then $V'=\lambda /2$ at its VEV. Substituting the solution (\ref{metricf}) into the motion equation (\ref{motion3}), one can obtain the cosmologic constant, i.e., \begin{eqnarray}\Lambda=(1+\ell)\frac{\kappa\lambda}{2\varrho}.\end{eqnarray}
One can see that the Lagrange-multiplier field $\lambda$ is constrained by the cosmological constant and it is not a new freedom.
Defining an effective  cosmological constant $\Lambda_e=\kappa\lambda/2\varrho$, one can see that $\Lambda=(1+\ell)\Lambda_e$. If $\ell\rightarrow0$, then $\Lambda_e\rightarrow\Lambda$. Then the new rotating BTZ-like black hole solution is
\begin{eqnarray}\label{metricb}
ds^2=-f(r)dt^2+\frac{(1+\ell)}{f(r)}dr^2+r^2(d\phi-\frac{j}{2r^2}dt)^2,\qquad f(r)= -m-(1+\ell)\Lambda_e r^2+\frac{j^2}{4r^2}.
\end{eqnarray}
This metric represents a purely radial LV black hole solution in a 3D spacetime. When $j\rightarrow0$, it is a static BTZ-like black hole solution
\begin{eqnarray}\label{}
 f(r)= -m-(1+\ell)\Lambda_e r^2,
\end{eqnarray}
which is consistent with the result of Eq. (29) in Ref. \cite{ding2023} when $D=3$.

Its Kretschmann scalar is
\begin{eqnarray}
R_{\mu\nu\rho\tau}R^{\mu\nu\rho\tau}=12\Lambda_e^2,
\end{eqnarray}
which is a finite constant in the whole spacetime as like as the original BTZ black hole, so there is also no curvature singularity at the origin. The  Kretschmann scalar of the BTZ black hole is $R_{\mu\nu\rho\tau}R^{\mu\nu\rho\tau}=12\Lambda^2$. In black hole chemistry \cite{kastor}, the negative cosmological constant $\Lambda_e$ can be thought of as a perfect fluid stress-energy with pressure $P=-\Lambda_e/8\pi$. This can be seen from its total energy-momentum tensor of the metric (\ref{metricb}),
\begin{equation}\label{totalmom}
\!\!\!\!\!\!\!\! T_\nu^\mu=\frac{1}{\kappa}\left(
\begin{array}{ccc}
-\epsilon  & 0  & 0 \\
0  & p_r &0    \\
0  & 0  & p_t
\end{array}
\right),
\end{equation}
where $\epsilon$ is the energy density, $p_r$ is the radial pressure and $p_t$ is the tangential pressure, which read
\begin{eqnarray}
-\epsilon=p_r=p_t=-\Lambda_e.
\end{eqnarray}
They are the same as the results of Eq. (49) in Ref. \cite{ding2023} when $D=3$. For the BTZ black hole, it is $-\epsilon=p_r=p_t=-\Lambda$. One can see that the radial and tangential pressures of the BTZ-like black hole spactime are identical, different from those in the four and higher dimensions \cite{maluf, ding2023}.

It has two horizons: inner(Cauchy) horizon $r_+$ and outer(event) horizon $r_-$, dependent on the coupling constant $\ell$, which can be read from $f(r_{\pm})=0$,
\begin{eqnarray}
r_{\pm}=\sqrt{\frac{2}{-(1+\ell)\Lambda_e}}\left(\sqrt{M
+\sqrt{-(1+\ell)\Lambda_e}J}\pm\sqrt{M-\sqrt{-(1+\ell)\Lambda_e}J}\right),
\end{eqnarray}
where $M=m/8$ is its ADM mass and $J=j/8$ is its angular momentum. Like Kerr black hole, it has an ergosphere $r_{erg}$ which can be read from the time-time component $g_{tt}=0$,
\begin{eqnarray}
r_{erg}=\sqrt{r_+^2+r_-^2}.
\end{eqnarray}
Its surface gravity of the event horizon $\hat{\kappa}$ is
\begin{eqnarray}\label{kappa}
\hat{\kappa}=-\frac{1}{\sqrt{1+\ell}}\Big((1+\ell)\Lambda_e r_++\frac{16J^2}{r_+^3}\Big),
\end{eqnarray}
which can be read from the formula \cite{ding2020},
\begin{eqnarray}
\hat{\kappa}=-\frac{1}{2}\lim_{r\rightarrow r_+}\sqrt{\frac{-1}{g_{rr}X}}\frac{dX}{dr}, \qquad X=g_{tt}-\frac{g_{t\phi}^2}{g_{\phi\phi}}.
\end{eqnarray}
It is easy to prove that in the present case, the function $X=-f(r)$.

\section{Thermodynamics and Central charges of the dual CFT}

According to the holographic theory \cite{witten}, one may expect that each sector of 3-dimensional  gravity which is either asymptotically anti-de Sitter(AdS) or AdS-like, there exists a dual 2-dimensional conformal field theory(CFT) that might live on the boundary of this AdS spacetime. Brown {\it et al} found a nontrivial central charge appears in the algebra of the canonical generators, which is just the Virasoro central charge \cite{brown}. From the black hole thermodynamics aspect, Yekta \cite{yekta} obtained the central charges of the CFT which can be constructed using the thermodynamics of the outer and inner horizons. He found the result was in complete agreement with that via the method of asymptotic symmetry group analysis \cite{liu1,liu2,liu3}. In this section, we study the central charges by using the thermodynamic method of black hole/CFT correspondence has been proposed in \cite{chen1,chen2,chen3}.

From the horizon equations $f(r_\pm)=0$, one can rewrite the black mass as \footnote{This mass also can be represented as $
M=PV_\pm+J\Omega_\pm/2,
$ where the term $PV_\pm$ is the work term, $J\Omega_\pm/2$ is its internal energy, then $M$ can be interpreted as its enthalpy.}
\begin{eqnarray}\label{}
M=-\frac{1}{8}(1+\ell)\Lambda_e r_\pm^2+\frac{2J^2}{r_\pm^2}.
\end{eqnarray}
The temperature of outer horizon $T_+$ is defined by (cf. Eq. \ref{kappa}),
\begin{eqnarray}\label{tp}
T_+=-\frac{1}{2\pi\sqrt{1+\ell}}\Big[(1+\ell)\Lambda_e r_++\frac{16J^2}{r_+^3}\Big]=-\frac{\sqrt{1+\ell}\Lambda_e}{2\pi r_+}(r_+^2-r_-^2).
\end{eqnarray}
The temperature of the inner horizon $T_-$ should be a geometrical positive quantity and is constant over the inner horizon \cite{castro}, so it should be,
\begin{eqnarray}\label{tm} T_-=\frac{1}{2\pi\sqrt{1+\ell}}\Big[(1+\ell)\Lambda_e r_-+\frac{16J^2}{r_-^3}\Big]=-\frac{\sqrt{1+\ell}\Lambda_e}{2\pi r_-}(r_+^2-r_-^2).
\end{eqnarray}
In Ref. \cite{kastor}, the negative cosmological constant $\Lambda_e$ can be thought of a pressure $P=-\Lambda_e/8\pi$, and the black mass $M$ can be taken as the enthalpy from classical thermodynamics, rather than as the total energy  of the spacetime. This enthalpy $M$ is required to both form a black hole and place it into its cosmological environment \cite{ding2016}.
 Then there are three thermodynamical quantities: entropy $S_\pm$ of the inner and outer horizons, angular momentum $J$ and pressure $P$, and mass $M$ can be expressed by the function of these quantities, i.e., $M=M(S,J,P)$. Their conjugate quantities are temperature $T$, angular velocity $\Omega$ and volume $V$, respectively. The entropy can be obtained by $S=A/4$, where $A$ is the horizon area\footnote{For the entropy-area relation $S=A/4$, see the appendix B for more details.}.

In Ref. \cite{ding2023}, we found that in this LV spacetime, the area and the volume should be redefined by
\begin{eqnarray}\label{spm}
&&A=\sqrt{1+\ell}\int r_h^{D-2}d\Omega_{D-2}=\sqrt{1+\ell}r_h^{D-2}\Omega_{D-2},\\
&& V=\sqrt{1+\ell}\int dr\sqrt{-g_D}d\Omega_{D-2}=(1+\ell)\frac{r_h^{D-1}}{D-1}\Omega_{D-2},
\end{eqnarray}
where $r_h$ is the radius of the horizon and $D$ is the number of dimension of the spacetime, and here $D=3$. So the results of the area and volume of the inner and outer horizons are, \footnote{Note that the above entropies and volumes can also be derived via the thermodynamic method  \begin{eqnarray}\label{}
S_\pm=\int\frac{dM}{T_\pm}=\frac{1}{2}\sqrt{1+\ell}\pi r_\pm,\qquad V_\pm=\left(\frac{\partial M}{\partial P}\right)_{S,J}=(1+\ell)\pi r_\pm^2.\nonumber
\end{eqnarray}}
\begin{eqnarray}\label{}
A_\pm=4S_\pm=\sqrt{1+\ell}\int^{2\pi}_0 r_\pm d\phi=2\sqrt{1+\ell}\pi r_\pm,\qquad V_\pm=\sqrt{1+\ell}\int^{2\pi}_0\int_0^{r_\pm} dr \sqrt{-g_3} d\phi=(1+\ell)\pi r_\pm^2,
\end{eqnarray}
where the three dimensional metric determinant $\sqrt{-g_3}=\sqrt{(1+\ell)}r$.
 The angular velocities $\Omega_\pm$ are,
\begin{eqnarray}\label{}
\Omega_\pm=\left(\frac{\partial M}{\partial J}\right)_{S,P}=\sqrt{8\pi (1+\ell)P}\frac{r_\mp}{r_\pm}.
\end{eqnarray}
It is easy to see that there exists the first law of thermodynamics and the Smarr formula for the outer horizon\footnote{What is the physical meaning about the term ``$0\cdot M$"? From the Eq. (39) of Komar integral in Ref. \cite{ding2023}, \begin{eqnarray}\label{}
\frac{D-2}{8\pi}\int_{\partial S_\infty}\left(\nabla^b\xi^a+\frac{2}{D-2}\Lambda_e\omega^{ab}\right)d\Sigma_{ab}
=-\frac{(D-3)M}{\sqrt{1+\ell}},\nonumber
\end{eqnarray} where $\partial S_{\infty}$ is the outer boundary at infinity, $\xi^a$ is the timelike Killing vector, $\omega^{ab}$ is the Killing potential, $d\Sigma_{ab}$ is the surface element of $\partial S_\infty$. When the dimension number $D=3$, the zero term is also obtained. Since this integral gives the total energy \cite{geroch,wald,jing1994} of the whole spacetime, one can see it means that the total energy of the three dimensional spacetime is zero.}
\begin{eqnarray}\label{}
dM= T_+ dS_++\Omega_+ dJ+V_+ dP,\qquad 0\cdot M= T_+ S_+-2PV_++\Omega_+ J.
\end{eqnarray}
For the inner horizon, the Killing vector is spacelike inside the event horizon, then one should assign negative energy $-M$ to the inner horizon \cite{castro}, similar to the negative energies within the ergosphere. So  the first law of thermodynamics and the Smarr formula for the inner  horizon are
\begin{eqnarray}\label{}
dM=- T_- dS_-+\Omega_- dJ+V_- dP,\qquad 0\cdot M=- T_- S_--2PV_-+\Omega_- J.
\end{eqnarray}

It is obviously that the equality $T_+S_+=T_-S_-$ is true from the relations (\ref{tp}), (\ref{tm}) and (\ref{spm}). This means that the entropy product of the inner and outer horizon
\begin{eqnarray}\label{}
S_+S_-=\sqrt{\frac{1+\ell}{-\Lambda_e}}\pi^2J,
\end{eqnarray}
is universal (mass-independent). So the central charges of the left- and right-moving sectors in the dual CFT must be the same,
 \begin{eqnarray}\label{}
c_L=c_R=6\frac{d}{dJ}\left(\frac{S_+S_-}{4\pi^2}\right)=\frac{3}{2}\sqrt{\frac{1+\ell}{-\Lambda_e}}.
\end{eqnarray} These results of 2-dimensional CFT on the boundary can be used to describe 3-dimensional AdS gravity in the bulk.

Defining the left-moving temperature $T_L$ and right-moving temperature $T_R$ of the dual CFT,
 \begin{eqnarray}\label{}
T_L=\left(\frac{1}{T_+}+\frac{1}{T_-}\right)^{-1}=-\frac{\sqrt{1+\ell}}{2\pi}\Lambda_e(r_+-r_-),\qquad T_R=\left(\frac{1}{T_+}-\frac{1}{T_-}\right)^{-1}=-\frac{\sqrt{1+\ell}}{2\pi}\Lambda_e(r_++r_-),
\end{eqnarray}
then the entropies of the inner and outer horizon can be represented by
 \begin{eqnarray}\label{}
S_\pm=\frac{\pi^2}{3\sqrt{-(1+\ell)\Lambda_e}}(c_LT_L\pm c_RT_R).
\end{eqnarray}
In addition, defining the energies of the left- and right-moving sectors of the dual CFT as
 \begin{eqnarray}\label{}
E_L=\frac{\pi^2}{6\sqrt{-(1+\ell)\Lambda_e}}c_LT_L^2=\frac{M-\sqrt{-(1+\ell)\Lambda_e}J}{2}, E_R=\frac{\pi^2}{6\sqrt{-(1+\ell)\Lambda_e}}c_RT_R^2=\frac{M+\sqrt{-(1+\ell)\Lambda_e}J}{2},
\end{eqnarray}
then the black hole mass $M$ and angular momentum $J$ can be represented by
 \begin{eqnarray}\label{}
M=E_L+E_R, \qquad J=\frac{1}{\sqrt{-(1+\ell)\Lambda_e}}(E_L-E_R).
\end{eqnarray}
In this way, we use the central charges on the boundary to reappear  the entropy, mass and angular momentum in the bulk.

\section{Summary}

In this paper, we have studied the black hole solution in Einstein gravity coupled to a bumblebee field in a 3-dimensional spacetime.
We find that there is no black hole solution with nonzero cosmological constant $\Lambda\neq0$ for the choice of the bumblebee potential $V=kx^2/2$. It exists only under the choice of a linear functional form with a Lagrange-multiplier field $\lambda$, $V=\lambda x/2$. Under this condition, we obtain an exactly rotating BTZ-like black hole solution with a negative cosmological constant and find that the additional field $\lambda$ is strictly constrained by the bumblebee field motion equation and can be absorbed in an effective cosmological constant $\Lambda_e$. When the angular momentum $j\rightarrow0$, it is a static BTZ-like black hole solution. The bumblebee field  affects the locations of the black hole horizon and ergosphere.

This black hole is different from 4-dimensional Schwarzschild-like \cite{casana} or Kerr-like \cite{ding2020,ding2021} black holes that it cannot be asymptotically flat and has no curvature singularity at the origin. It is asymptotically AdS and have isotropic pressures, that is different from the 4- or higher dimensional AdS-like black holes \cite{maluf,ding2023} whose pressures are anisotropic.

Then we study its thermodynamics, find that the concepts of the area and volume of the horizons in this LV spacetime should be renewed due to the nontrivial contribution of coupling between the bumblebee field $B^\mu$ and the Ricci tensor $R_{\mu\nu}$. Only in this way, the entropy-area relation $S=A/4$, the first law and the Smarr formula can still be constructed. We also study the AdS/CFT correspondence of this black hole, find
that the entropy product of its inner and outer horizons is universal. So the central charges of the
dual CFT on the boundary can be obtained via the thermodynamic method, and they can reappear
black hole mass and angular momentum in the bulk.

\begin{acknowledgments} This work was supported by the Scientific Research Fund of the Hunan Provincial Education Department under No. 19A257, the National Natural Science Foundation (NNSFC)
of China (grant No. 11247013), Hunan Provincial Natural Science Foundation of China (grant No. 2015JJ2085 and No. 2020JJ4284).
\end{acknowledgments}

\appendix\section{Some quantities I}
In this appendix, we showed the nonezero components of Einstein's tensor for the metric (\ref{metric}). They are as following
\begin{eqnarray}
&&G_{00}=e^{-2\psi}r^2\Omega^2(\Phi''+\Phi'^2-\Phi'\psi')
+e^{-2\psi}r^2\Omega(-\Omega''+\Phi'\Omega'+\Omega'\psi')\nonumber\\&& \qquad\qquad-\frac{1}{4}e^{-2\Phi-2\psi}r^2(e^{2\Phi}+3r^2\Omega^2)\Omega'^2-3e^{2\psi}r\Omega\Omega'
+e^{2\Phi-2\psi}\frac{\psi'}{r},\\
&&G_{02}=e^{-2\psi}r^2\Omega(\Phi''+\Phi'^2-\Phi'\psi')
+\frac{1}{2}e^{-2\psi}r^2(-\Omega''+\Phi'\Omega'-\Omega'\psi')\nonumber\\&& \qquad\qquad-\frac{3}{4}e^{-2\Phi-2\psi}r^4\Omega\Omega'^2-\frac{3}{2}e^{-2\psi}r\Omega',\\
&&G_{11}=\frac{\Phi'}{r}+\frac{1}{4}e^{-2\Phi}r^2\Omega'^2,\\
&&R_{11}=\frac{\psi'}{r}-(\Phi''+\Phi'^2-\Phi'\psi')+\frac{1}{2}e^{-2\Phi}r^2\Omega'^2,\\
&&G_{22}=e^{-2\psi}r^2(\Phi''+\Phi'^2-\Phi'\psi')-\frac{3}{4}e^{-2\Phi-2\psi}r^4\Omega'^2.
\end{eqnarray}
$\bar B_{\mu\nu}$  are
\begin{eqnarray}
&&\bar B_{00}=-\frac{\varrho b_0^2}{2}e^{-2\psi}(e^{-2\Phi}+r^2\Omega^2)(\Phi''+\Phi'^2-\Phi'\psi')+\varrho b_0^2e^{-2\psi}r^2\Omega(\Omega''-\Phi'\Omega'-\Omega'\psi')\nonumber\\&& \qquad\qquad+\frac{\varrho b_0^2}{2}e^{-2\Phi-2\psi}r^2(e^{2\Phi}+r^2\Omega^2)\Omega'^2+3\varrho b_0^2e^{-2\psi}r\Omega\Omega'-\frac{\varrho b_0^2}{2r}e^{-2\psi}(e^{2\Phi}+r^2\Omega^2)\psi',\\
&&\bar B_{02}=-\frac{\varrho b_0^2}{2}e^{-2\psi}r^2\Omega(\Phi''+\Phi'^2-\Phi'\psi')+\frac{\varrho b_0^2}{2}e^{-2\psi}r^2(\Omega''-\Phi'\Omega'-\Omega'\psi')\nonumber\\&& \qquad\qquad +\frac{\varrho b_0^2}{2}e^{-2\Phi-2\psi}r^4\Omega\Omega'^2-\frac{\varrho b_0^2}{2}e^{-2\psi}r(\Omega\psi'-3\Omega'),\\
&&\bar B_{11}=-\frac{\varrho b_0^2}{2}(\Phi''+\Phi'^2-\Phi'\psi')-\frac{\varrho b_0^2}{2r}(2\Phi'-\psi'),\\
&&\bar B_{22}=-\frac{\varrho b_0^2}{2}e^{-2\psi}r^2(\Phi''+\Phi'^2-\Phi'\psi')+\frac{\varrho b_0^2}{2}e^{-2\Phi-2\psi}r^4\Omega'^2-\frac{\varrho b_0^2}{2}e^{-2\psi}r\psi'.
\end{eqnarray}

\section{Details of the entropy-area relation}
Since the bumblebee theory has a non-minimall coupling between the vector field $B_\mu$ and the Ricci tensor $R_{\mu\nu}$, is the entropy-area relation  still $S=A/4$? This coupling contribution is not trivial, but the above entropy-area still holds as long as the redefinition of the area. We derive it in the following via the analysis of the Euclidean action for the gravitation field.

The Euclidean on-shell action for the rotating BTZ-like Einstein-bumblebee black hole is \cite{kallosh,jing1996}
 \begin{eqnarray}\label{actionE}
I=\frac{\sqrt{1+\ell}}{16\pi}\int_\Sigma(\mathcal{L}_{matter}-R)\sqrt{g}d^3x
+\frac{\sqrt{1+\ell}}{8\pi}\int_{\partial\Sigma}(K-K_0)\sqrt{h}d^2x,
\end{eqnarray}
with
 \begin{eqnarray}\label{}
\mathcal{L}_{matter}=2\Lambda-\varrho B^\mu B^\nu R_{\mu\nu}+16\pi\Big[\frac{1}{4}B^{\mu\nu}B_{\mu\nu}+\frac{\lambda}{2}(B^\mu B_\mu-b^2)\Big],
\end{eqnarray}
where $\Sigma$ is the spacetime region with the metric $g$, $K$ is the trace of the extrinsic curvature of the boundary $\partial\Sigma$, $K_0$ is that when $K$ is for the flat spacetime, $h$ is the determinant of the induced metric on $\partial\Sigma$, $g$ is the Euclidean black hole metric, which can be gotten by setting $t\rightarrow i\tau$ in Eq. (\ref{metricb}) \footnote{In Ref. \cite{jing1996}, $\varphi'\rightarrow i\varphi$ is also set. We find it is wrong duo to that the Euclidean black hole metric determinant $g$ will be negative.},
 \begin{eqnarray}\label{metricE}
ds^2=\Big[f(r)-\frac{16J^2}{r^2}\Big]d\tau^2+\frac{1+\ell}{f(r)}dr^2-8Jid\tau d\phi+r^2d\phi^2,\qquad f(r)=-8M-(1+\ell)\Lambda_er^2+\frac{16J^2}{r^2},
\end{eqnarray}
where $\tau$ has period $\beta=2\pi/\hat\kappa$, and $\hat\kappa$ is given by the Eq. (\ref{kappa}).
Why the action (\ref{actionE}) is multiplied by the factor $\sqrt{1+\ell}$? Because in this LV spacetime, the volume and area elements should be redefined due to the nontrivial contribution of coupling between the bumblebee field $B^\mu$ and the Ricci tensor $R_{\mu\nu}$, cf. Eqs. (\ref{spm})\footnote{If the pre-factor $\sqrt{1+\ell}$ is missing, it will lets the entropy-area relation to $S=\sqrt{1+\ell}A/4$. }. In Refs. \cite{jing1994,kallosh}, the authors found that the entropy is
 \begin{eqnarray}\label{}
S=-\frac{\sqrt{1+\ell}}{8\pi}\int_{\partial \Sigma_h}(K-K_0)\sqrt{h}d^2x,
\end{eqnarray}
where $\partial\Sigma_h$ is the horizon.

In the Eq. (\ref{actionE}), the extrinsic curvature $K$ of the boundary $\partial\Sigma$ can be obtained by a spacelike unit normal vector $n_\mu=(0,\sqrt{g_{rr}},0)$ as following,
 \begin{eqnarray}\label{}
K=h^{\mu\nu}\nabla_\mu n_\nu,\qquad h_{\mu\nu}=g_{\mu\nu}-n_{\mu}n_{\nu},
\end{eqnarray}
where $h_{\mu\nu}$ is the induced metric on the hypersurface $\partial \Sigma$. Using the metric in Eq. (\ref{metricE}), we find that,
 \begin{eqnarray}\label{}
K=\frac{1}{2\sqrt{g_{rr}}Y}[g_{\phi\phi}\partial_rg_{\tau\tau}-2g_{\tau\phi}\partial_rg_{\tau\phi}
+g_{\tau\tau}\partial_rg_{\phi\phi}],\qquad \sqrt{h}=\sqrt{Y},\qquad Y=g_{\tau\tau}g_{\phi\phi}-g_{\tau\phi}^2.
\end{eqnarray}
Then,
 \begin{eqnarray}\label{}
K=\frac{1}{\sqrt{(1+\ell)f(r)}}\Big[\frac{1}{2}\partial_r f(r)+\frac{1}{r}f(r)\Big],\qquad \sqrt{h}=r\sqrt{f(r)},\qquad K_0=\frac{1}{r}.
\end{eqnarray}
The second integration in Eq. (\ref{actionE}) becomes,
 \begin{eqnarray}\label{}
-\frac{\sqrt{1+\ell}}{8\pi}\int_{\partial\Sigma}(K-K_0)\sqrt{h}d^2x=-\frac{1}{\hat\kappa}\frac{A(r)}{4}
\Big[\frac{\partial_r f}{2\sqrt{1+\ell}}+\frac{1}{r}\big(\frac{ f}{\sqrt{1+\ell}}-\sqrt{f}\big)\Big],
\end{eqnarray}
where $A(r)=2\pi\sqrt{1+\ell}r$. Applying it on the horizon, the entropy is,
 \begin{eqnarray}\label{}
S=-\frac{\sqrt{1+\ell}}{8\pi}\int_{\partial \Sigma_h}(K-K_0)\sqrt{h}d^2x=\frac{A(r_h)}{4}.
\end{eqnarray}
Therefore, one can see that, as long as the area is redefined due to the nontrivial contribution of coupling between the bumblebee field $B^\mu$ and the Ricci tensor $R_{\mu\nu}$, the entropy-area relation $S=A/4$ is still hold in the bumblebee LV spacetime.

\end{document}